\documentclass[fleqn,twoside]{article}
\usepackage[headings]{espcrc2}

\readRCS
$Id: espcrc2.tex,v 1.2 2004/02/24 11:22:11 spepping Exp $
\usepackage{graphicx}
\usepackage[figuresright]{rotating}

\newcommand{\AmS}{{\protect\the\textfont2
  A\kern-.1667em\lower.5ex\hbox{M}\kern-.125emS}}

\title{New one-loop techniques and first applications to LHC phenomenology}

\author{\underline{T. Binoth}\address[Edi]{School of Physics, 
        SUPA, The University of Edinburgh, \\ 
        Mayfield Road, Edinburgh EH9 3JZ, Scotland, UK.}, %
        A. Guffanti\addressmark[Edi],
	J.-Ph. Guillet\address[LAPTH]{Laboratoire d'Annecy-le-Vieux de Physique Th\'eorique 
                    LAPTH, \\ B.P. 110, F-74941 Annecy-le-Vieux Cedex, France.}, %
        S. Karg\address[Wue]{Faculty for Theoretical Physics, University of Wuerzburg, \\
        Am Hubland, D-97074 Wuerzburg, Germany.}, %
        N. Kauer\addressmark[Wue] and
	T. Reiter\addressmark[Edi]}
       
\runtitle{}
\runauthor{T. Binoth et al.}

\begin{document}


\begin{abstract}
In this talk we describe our approach for the computation of
multi-leg one-loop amplitudes and present some first results
relevant for LHC phenomenology. 
\vspace{1pc}
\end{abstract}

\maketitle

\section{Motivation}

\unitlength=1mm
\begin{picture}(0,0)
\put(120,110){ \parbox{3cm}{Edinburgh 2006/13\\
LAPTH-1154/06\\
June 2006} }
\end{picture}

In the next year the Large Hadron Collider at CERN will start
operating at the up to now unreached TeV energy scale. 
Apart from putting our theoretical understanding of the electroweak
symmetry breaking mechanism under scrutiny, new effects beyond the Standard
Model that occur in this energy region can also be explored 
\cite{Buttar:2006zd,Buscher:2005re}.  A successful quantitative analysis 
and comparison of the LHC data with
theoretical predictions will crucially rely not only on 
precise predictions for a variety of signal cross sections, but also
on a sufficiently precise knowledge of the various backgrounds.
Although in some favourable cases the Standard Model backgrounds
can be measured directly, for other important search channels this is not 
possible and backgrounds have to be determined through theoretical predictions.
This is especially worrisome, as QCD predictions for multi-particle
processes are  plagued
by large uncertainties arising from high powers of $\alpha_s$. 
Large scale uncertainties make it mandatory to include radiative
corrections beyond the leading order in QCD.  
Unfortunately, due to the complicated singularity structure of QCD and
the general combinatorial complexity of multi-scale problems,
one-loop computations  with more than four external partons/particles
are a highly non-trivial and time-consuming task.
Fortunately, at the dawn of the LHC era, this research direction is 
very actively pursued by multiple groups
\cite{Denner:2005nn,Denner:2005fg,Ellis:2005zh,Ellis:2006ss,Binoth:2005ff,vanHameren:2005ed,Giele:2004iy,Nagy:2003qn,Binoth:2002xh,Duplancic:2003tv,Ferroglia:2002mz,Binoth:1999sp}.

In this talk I present our approach to this problem. The aim of our
collaboration is to build a flexible and reliable tool, which allows for 
the evaluation of one-loop multi-leg amplitudes. It is based on a 
combination of algebraic and numerical methods.
The corresponding project is
called GOLEM (General One-Loop Evaluator for Matrix elements) and will
be discussed in the next section. Section 3 is dedicated to first applications
of these methods relevant for LHC phenomenology.    

\section{The GOLEM project}

A General One-Loop Evaluator for Matrix elements (GOLEM) should proceed from
a Feynman diagrammatic representation of a given scattering amplitude to 
a computer code which provides a numerically stable and accurate  answer 
for the desired cross section.

In a first step each amplitude can be represented in terms
of irreducible Lorentz tensors which correspond uniquely to gauge invariant 
operators: 
\begin{eqnarray}
\mathcal{A}(|p_j \rangle,{\epsilon^{\lambda}_j},\dots) = 
       \sum\limits_{I} \mathcal{A}_I(|p_j \rangle,{\epsilon^{\lambda}_j},\dots)  \,\, .\nonumber
\end{eqnarray}
The mapping of the diagrammatic input onto such a tensorial basis
can be accomplished with algebraic manipulation programs. 
We generally use FORM 3.1 \cite{Vermaseren:2000nd}.
Each Feynman diagram $\mathcal{G}_G$ can be projected onto these sub-amplitudes: 
\begin{eqnarray}
    \mathcal{A}(|p_j \rangle,{\epsilon^{\lambda}_j},\dots) = 
         \sum\limits_{G} \mathcal{G}_{G}(|p_j \rangle,{\epsilon^{\lambda}_j},\dots) && \nonumber\\
     = \sum\limits_{I} \sum\limits_{G}  \mathcal{C}_{IG}(s_{jk})\,
                                       \mathcal{T}_{I}(|p_j \rangle,{\epsilon^{\lambda}_j},\dots)  && . \nonumber
\end{eqnarray}
The coefficients $\mathcal{C}$ of the gauge invariant operators are momentum 
space integrals where free Lorentz indices are contracted with
external momentum vectors. In a third step these coefficients, which depend on Mandelstam variables $s_{ij}=(p_i+p_j)^2$ only,
are expressed in terms of some integral basis to be discussed below:
\begin{eqnarray}
    \mathcal{A}(|p_j \rangle,{\epsilon^{\lambda}_j},\dots) 
     = \sum\limits_{BIG}   \mathcal{C}_{BIG}(s_{jk},\dots) && \nonumber\\ \times \, I_B\,
                                       \mathcal{T}_{I}(|p_j \rangle,{\epsilon^{\lambda}_j},\dots) && . \nonumber  
    \end{eqnarray}
In a forth step, the coefficients   
have to be exported to some numerical programming language.
Optionally they can be simplified beforehand using  
algebraic programs like MAPLE or MATHEMATICA:
 \begin{eqnarray}
    \mathcal{A}(|p_j \rangle,{\epsilon^{\lambda}_j},\dots) 
     = \sum\limits_{BIG} \, I_B \,\mathcal{T}_{I}(|p_j \rangle,{\epsilon^{\lambda}_j},\dots) &&  \nonumber\\
     \times \,\textrm{simplify[} \mathcal{C}_{BIG}(s_{jk},\dots)\,] && .
                                          \nonumber
    \end{eqnarray}   

We now discuss the integral basis used for our reduction algorithms.
Following \cite{Davydychev:1991va} one can represent any given tensor integral
by  Feynman parameter integrals in shifted dimensions:
\begin{eqnarray}
       I_{N}^{\mu_1\dots \mu_R} = \sum \tau^{\mu_1\dots \mu_R}(r_{j_1},\dots,r_{j_r},g^m) &&\nonumber\\
                                \times  I_{N}^{n+2m}(j_1,\dots,j_r) && . \nonumber  
   \end{eqnarray}
By applying differentiation by parts in parameter space and 
$d=4$ kinematical identities \cite{Binoth:2005ff} one can map all these integrals
to the following basis without the need for higher dimensional scalar integrals
with $N>4$:
\begin{eqnarray}\label{golem_basis}
I^{n}_3(j_1, \ldots ,j_r) = -\Gamma \left(3-\frac{n}{2} \right)  && \nonumber\\ \times
\int_{0}^{1} \prod_{i=1}^{3} \, d z_i \,\frac{z_{j_1} \ldots z_{j_r}\delta(1-\sum_{l=1}^{3} z_l) }{ (-\frac{1}{2}\, z \cdot \mathcal{S}\cdot z)^{3-n/2}} \ \ ,\nonumber\\
I^{n+2}_3(j_1) = 
-\Gamma \left(2-\frac{n}{2} \right) && \nonumber\\  \times
\int_{0}^{1} \prod_{i=1}^{3} \, d z_i \, \frac{z_{j_1}\,\delta(1-\sum_{l=1}^{3} z_l)}{(-\frac{1}{2}\, z \cdot \mathcal{S} \cdot z)^{2-n/2}} \ \ ,\nonumber\\
I^{n+2}_4(j_1, \ldots ,j_r) = \Gamma \left(3-\frac{n}{2} \right) && \nonumber\\  \times
\int_{0}^{1} \prod_{i=1}^{4} \, d z_i  \, \frac{z_{j_1} \ldots z_{j_r}\delta(1-\sum_{l=1}^{4} z_l)}{(-\frac{1}{2}\, z \cdot \mathcal{S}\cdot z)^{3-n/2}} \ \ ,\nonumber \\
I^{n+4}_4(j_1) = \Gamma \left(2-\frac{n}{2} \right) && \nonumber\\ \times
\int_{0}^{1} \prod_{i=1}^{4} \, d z_i \, \frac{z_{j_1}\delta(1-\sum_{l=1}^{4} z_l)}{(-\frac{1}{2}\, z \cdot \mathcal{S}\cdot z)^{2-n/2}}
\end{eqnarray}
and scalar integrals $I^{n}_2$, $I^{n}_3$, $I^{n+2}_3$, $I^{n+2}_4$. 
We call this basis the GOLEM basis. Note that any $N$-point
function can be mapped to this basis without introducing 
inverse Gram determinants, i.e. $1/\det(G)^r$ with $G_{ij}=2\, r_i\cdot r_j$, $r_j=p_1+\dots+p_j$, 
which occur in Passarino-Veltman reduction.
There are three alternatives for evaluating the basis elements:
\begin{itemize}
\item[1.] algebraic reduction to the one-loop master integrals $I^{n}_2$, $I^{n}_3$, $I^{n+2}_4$,
\item[2.] semi-numerical reduction to scalar \\ integrals,
\item[3.] direct numerical evaluation. 
\end{itemize}
In this context, semi-numerical means that the same reduction formulas 
as in 1. are applied at the numerical level. 
In case 1. and 2.  Gram determinants are reintroduced.
The direct numerical evaluation is chosen whenever numerically
critical phase space regions are approached.

For the numerical evaluation of the GOLEM basis
integrals two different methods have been developed.
The first is based on contour deformation in Feynman parameter space \cite{Binoth:2005ff}
and the second on performing the Cauchy integration explicitly \cite{Binoth:2002xh}.
The latter method is faster, since (at most) 2- rather than 3-dimensional representations are used.
We have tested the robustness of the numerical routines for large
samples of realistic phase space points. 

The algebraic/numerical algorithms  are implemented
in the flexible Fortran 90 code \texttt{GOLEM90}. 

\section{Applications to LHC phenomenology}

Using the algorithms discussed above, several computations relevant for LHC
phenomenology have been carried out so far. We present three examples
here, others can be found in \cite{Binoth:2001vm,Binoth:2002qh,Binoth:2003xk}.

\subsection{The process $gg\to W^*W^* \to l\bar{\nu}\, \bar{l}'\nu'$}  

\begin{figure}[h]
\unitlength=1mm
\begin{picture}(80,55)
\put(0,0){\includegraphics[width=5.5cm,angle=90]{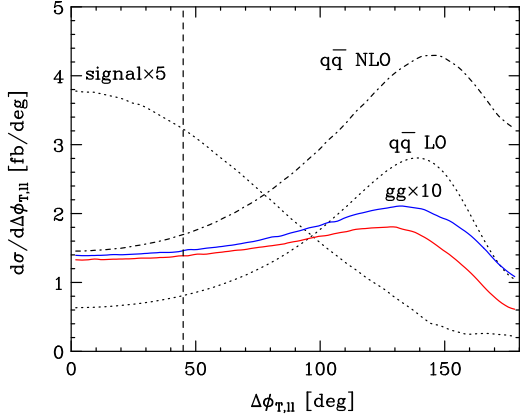}} 
\end{picture}
\caption{Azimuthal opening angle distribution for the charged lepton pair at the LHC.
The upper (lower) solid line is the gluonic contribution $(\times 10)$ including (excluding)
the massive third family quarks. The other curves are the Higgs signal $(\times 5)$ and
LO/NLO quark contribution (MCFM). 
For details see \cite{Binoth:2005ua}.}
\label{ggww}
\end{figure}

This process contributes to the dominant irreducible background to the Higgs production process 
$gg\to H \to W^*W^*$.  It is also relevant below the $WW$ threshold.
The gluon-vector boson interaction is mediated by a quark loop.
On-shell production results with massless \cite{Glover:1988fe} and massive  quarks 
\cite{Kao:1990tt} have been known for a long time. For on-shell $W$ bosons, a
phenomenological analysis can be found in \cite{Duhrssen:2005bz}. 
We have now a full calculation for off-shell $W$-pairs, where 
the first two quark generations are treated massless, but the bottom and top masses
in the third generation are kept. In the latter case the amplitude contains six
different scales: the Mandelstam variables $s$, $t$, the $W$-virtualities
$s_3$, $s_4$ and the quark masses $m_b$, $m_t$.
For this amplitude, it is possible to reduce algebraically down to master integrals, 
$I_2^n$, $I_3^n$, $I_4^{n+2}$.
The amplitude is represented by nine independent gauge-invariant structures.
Each coefficient can be written as a linear combination of
27 basis functions. The algebraic simplification results in 
amplitude representations which have at most one inverse Gram determinant.
The Gram determinant is related to the
transverse momentum of the $W$ bosons by $p_T^2(W)=\det(G)/s^2$. 
In this computation there is a tiny phase space region
$p_T(W)<0.1$ GeV, $|s_{3,4}-M_W^2| \gg M_W\Gamma_W$ which is
numerically problematic, but it does not contribute sizably to
the relevant cross sections \cite{Binoth:2005ua}.    
In Tab.~\ref{tab:gg2ww} we show typical LHC cross sections for 
two massless generations and including the third generation. We also compare to the 
full NLO result for $q\bar q\to W^*W^*$, which is computed 
using MCFM \cite{Campbell:2002tg}.
Cross sections without selection cuts (tot), with standard LHC cuts (std)
and Higgs search cuts (bkg) are given (see \cite{Binoth:2005ua} for details).
\begin{table}
\begin{tabular}{|c|c|c|c|}
\hline
  &  $gg$ (2 gen.)  & $gg$ (3 gen.)  &   $\frac{\sigma_{NLO}+\sigma_{gg}}{\sigma_{NLO}}$  \\
\hline
$\sigma_{tot}$ & 53.61(2)  & 60.12(7) &  1.04 \\
\hline
$\sigma_{std}$ & 25.89(1)  & 29.79(2) &  1.06 \\
\hline
$\sigma_{bkg}$ & 1.385(1)  & 1.416(3) &  1.30 \\
\hline
\end{tabular}
 \caption{\label{tab:gg2ww} Background cross sections in fb for  charged lepton pair production in gluon fusion
$gg\to W^*W^*\to l\bar{\nu}\,\bar{l}'\nu'$. Results for two and three
quark generations are shown for different selection cuts. The last column 
shows the importance of the gluon fusion process (3 gen.) relative to the
quark induced channel in NLO QCD.}
\end{table}
This calculation shows that for Higgs search cuts the $gg\to W^*W^*$ process 
enhances the previously known $W$-pair background by 
approximately 30\%
and has to be taken into account in experimental studies.
In Fig.~\ref{ggww}, we show the charged-lepton azimuthal opening angle distribution. 

\subsection{The process $gg\to HHH$}

Multi-Higgs production via gluon fusion allows in principle 
to measure the triple and quartic Higgs self-coupling. 
A full calculation
of three Higgs boson production in gluon fusion was accomplished  recently
by \cite{Plehn:2005nk}. Four topologies with different
sensitivities can be distinguished.
\unitlength=1mm
\begin{picture}(85,40)
\put(4,18){\includegraphics[width=3.cm]{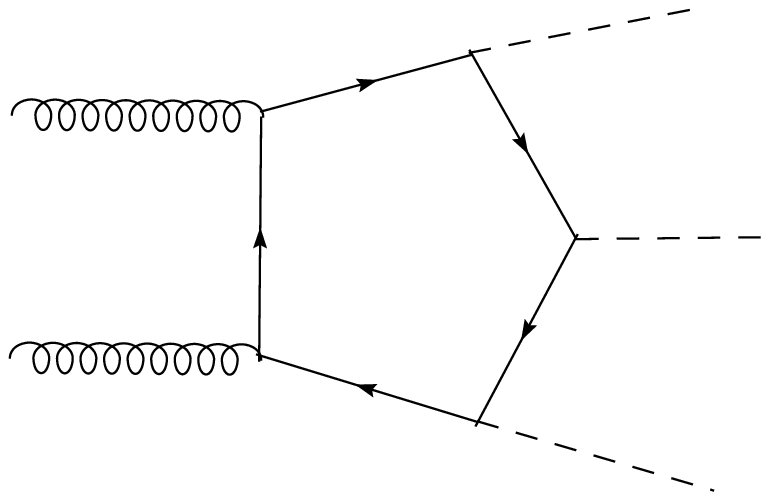}}
\put(41,18){\includegraphics[width=3.cm]{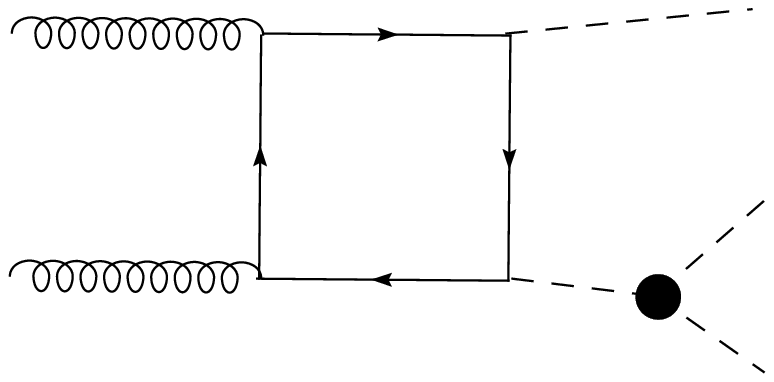}}
\put(4,2){\includegraphics[width=3.cm]{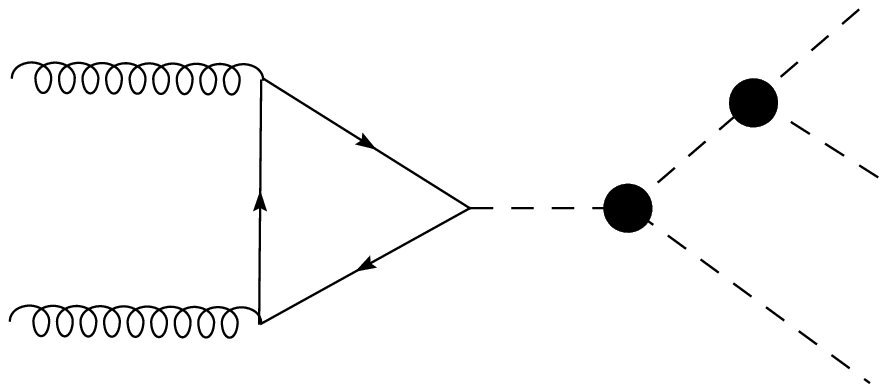}}
\put(41,2){\includegraphics[width=3.cm]{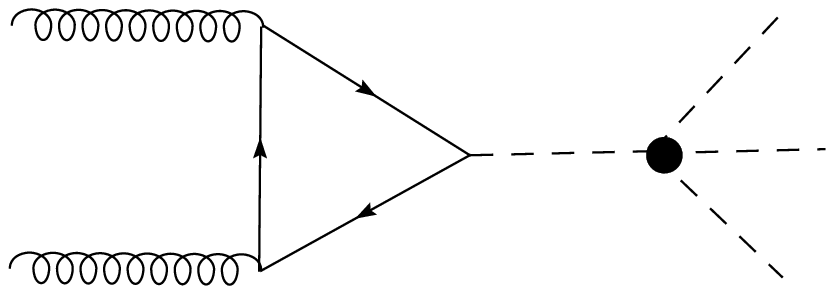}}
\end{picture}

\begin{figure}[h]
\unitlength=1mm
\begin{picture}(80,45)
\put(0,-8){\includegraphics[width=7.2cm]{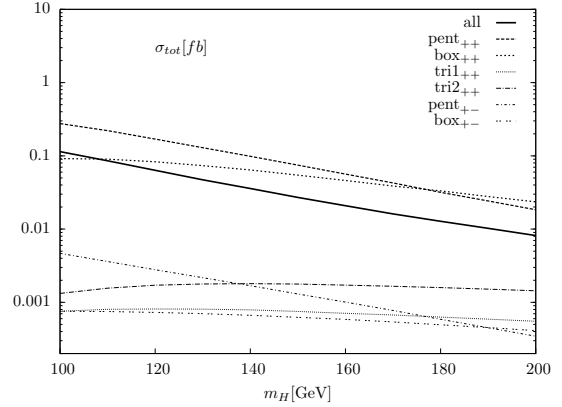}} 
\end{picture}
\caption{The total triple Higgs production cross section vs. $m_H$
for different topologies (tri1 $\sim \lambda_{4H}$, tri2 $\sim \lambda_{3H}^2$).
The $+-$ helicity component and all triangle topologies are suppressed.
}
\label{fig:gghhh}
\end{figure}

The amplitude can be decomposed into five independent 
gauge invariant structures $\textrm{tr}(\mathcal{F}_1\mathcal{F}_2)$,
$p_2\cdot \mathcal{F}_1\cdot p_l\; p_1\cdot \mathcal{F}_2\cdot p_j$, $l,j\in \{3,4\}$,
where $\mathcal{F}$ is the gluon field strength tensor. 
Some of the respective
coefficients are related by Bose symmetry. Evaluating all
coefficients and testing these relations serves as
a check of the calculation.
Algebraic reduction to 68 master integrals is possible
and one finds a numerically stable representation of the amplitude
with at most one inverse Gram determinant. The result is summarised
in Fig. \ref{fig:gghhh}.
Destructive interference effects are present between
pentagon box and box triangle topologies. The total cross section for
$m_H=120$ GeV, being only of the order of 0.06 fb, is out of reach  at the LHC. 
Note that in Beyond Standard Model scenarios
with an enlarged fermion or Higgs sector this cross section could be much 
larger. We  compared our full result to the heavy top 
limit and observed that for three Higgs boson production
the $m_{top} \to \infty$ approximation is not applicable.
Here, the kinematically favoured region is above the $2\,m_{top}$ threshold,
rather than below, as required by the heavy top limit.
 
\subsection{The $PP\to 4\; jet$ amplitude}

Up to now no full LHC process with $2\to 4$ kinematics has
been evaluated. To have a good theoretical prediction for
the huge four jet rate, next-to-leading order corrections
to the partonic reactions $gg\to gggg$, $gg\to ggq \bar q$, 
$gg\to q \bar q q \bar q$, $q \bar q \to q \bar q q \bar q$  
plus crossings need to be evaluated. 
By using our GOLEM algorithm we have produced a 
representation of the six quark amplitude in terms of 
the GOLEM basis integrals (\ref{golem_basis}) which allows for the
semi-numerical evaluation of this amplitude. The evaluation time
of one kinematical point is of the order of a second.
Alternatively we have produced an algebraic representation of 
the amplitude in terms of master integrals $I_2^n$, $I_3^n$ and
$I_4^{n+2}$. The  two independent approaches allow
for an efficient debugging of our algebraic/numerical procedures.
As this work is not yet completed we just illustrate this by providing 
a numerical evaluation of the most complicated Feynman diagram 
in this calculation, the 6-quark pentagon diagram
\begin{picture}(100,24)
\put(30,1){\includegraphics[width=2.0cm]{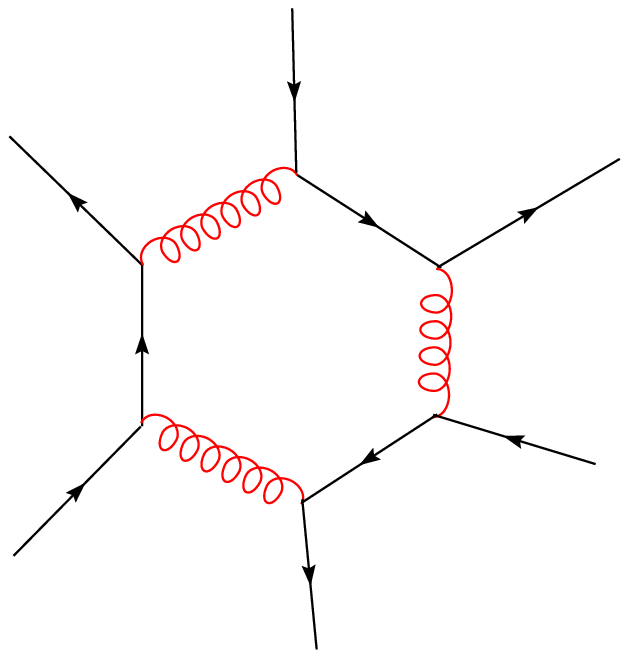}}
\end{picture}

This diagram can be written as
\begin{eqnarray}
&&A^{\lambda_1\lambda_2\lambda_3\lambda_4\lambda_5\lambda_6}(k_1,\ldots,k_6)
= 
 \nonumber \\&&
\frac{g_s^6}{(4\pi)^{2}} \frac{1}{s}[ 
\frac{A}{\epsilon^2} + 
\frac{B}{\epsilon}+ C + {\cal O}(\epsilon)]  \nonumber
\end{eqnarray}
For the helicity amplitude ++++++ and  the kinematical point

\begin{tabular}{|l|rrrr|}
\hline
$k$ & $k^0$ & $k^1$ & $k^2$ & $k^3$ \\
\hline
$k_1$ & 7000.     &    0.      &      0.     &  7000.\\
$k_2$ & 7000.     &    0.      &      0.     & -7000.\\
$k_3$ & 5105.37& 1752.05& -972.19 & -4695.74\\ 
$k_4$ & 4582.45& -3465.91 &  1373.64  & 2664.47\\
$k_5$ & 2917.90& 2182.21 &  876.20  &1727.53\\
$k_6$ & 1394.27& -468.35 & -1277.65  &  303.74\\
\hline
\end{tabular}

\smallskip

we find, up to an overall phase,
\begin{eqnarray}
A &=& 3.27888 + \, i\, 0.950546 \nonumber\,\, ,\\
B &=& 10.7288 + \, i\,  15.53100\nonumber\,\, ,\\
C &=& -1.16693 + \, i\,  59.1171\nonumber \,\, .
\end{eqnarray}
The evaluation time for this point with our code is about 0.2 sec on a standard PC (Pentium 4, 2.8 GHz).
The evaluation of the full 6-quark amplitude including real emission corrections
is in progress.

\section{Conclusion}

In this talk we presented our GOLEM approach for the 
evaluation of one-loop multi-leg processes relevant for
precise LHC phenomenology. The formalism allows in principle for the
evaluation of general $N$-point processes with massive and/or massless
particles.
The algebraic/numerical reduction and evaluation techniques   
are implemented in the Fortran 90 code \texttt{GOLEM90}.
As applications, recent results for the gluon-induced processes 
$gg\to W^*W^* \to l\bar{\nu}\, \bar{l}'\nu'$, $gg\to HHH$ and the hexagon process
$q\bar q\to q\bar qq\bar q$ were presented.
\section*{Acknowledgement}

T.B. would like to thank the conference organizers for the stimulating
conference in Eisenach.
The work of T.B., S.K. and N.K. was supported  by the Deutsche
Forschungsgemeinschaft (DFG) under the grant number Bi1050/1-1.\&2.
and the Bundesministerium fuer Bildung und Forschung (BMBF) under contract number
05HT1WWA2.

\end{document}